\documentclass [12pt,preprint]{aastex}
\begin{document}

\title{Cosmic Star Formation History from Local Observations and an Outline 
for Galaxy Formation and Evolution}

\author{F.D.A. Hartwick}

\affil{Department of Physics and Astronomy, \linebreak University of Victoria, 
Victoria, BC, Canada, V8W 3P6}
\begin {abstract}
The goal of this investigation is to reconstruct the cosmic star formation 
rate density history from local observations and in doing so to gain insight 
into how galaxies might have formed and evolved. A new chemical evolution model
is described which accounts for the formation of globular clusters 
as well as the accompanying field stars. When this model is used in conjunction
with the observed age metallicity relations for the clusters and with input 
which allows for the formation of the nearly universally observed 
bimodal distribution of globular clusters, star formation rates are obtained.
By confining attention to a representative volume of the local universe, these
rates allow a successful reconstruction of the Madau plot while complementary 
results similtaneously satisfy many local cosmological constraints. A physical
framework for galaxy formation is presented which incorporates the results from
this chemical evolution model and assumes an anisotropic collapse. In addition
to providing the `classical' halo, bulge and disk components, the model also 
predicts a new stellar halo component with peak [Fe/H]$\sim-0.8$ and disk-like 
angular momentum and allows for the formation of a thick disk as outlined by 
the group of metal rich globular clusters. Milky Way counterparts of the latter
two components are identified.
\end{abstract}

\keywords{Galaxy: formation --- Galaxy: halo ---  globular clusters: general}

\section {Introduction}

Observations of the luminosity functions of high redshift galaxies enable one
to determine in situ the star formation rate density (SFRD) as a function of 
redshift. A plot of these two quantities has become known as a Madau diagram 
(Lilly et al.\ 1996, Madau et al.\ 1996, 1998). Locally, we observe the results
of this early evolution in the form of old stars and globular clusters. Hence, 
given a model of chemical evolution from which the metallicity distribution 
functions (MDF's) are obtained, combined with the corresponding 
age-metallicity relations, we should be able to reconstruct the star formation
history from these local observations and compare it to what is currently 
observed at high redshift. 

A very preliminary investigation of this
problem along these lines was given by Hartwick (1999). Since then,
many more high redshift observations have become available (i.e. the Madau 
plot has been extended), the cosmological parameters have become better 
defined (providing a more reliable age-redshift relation), and age-metallicity 
relations for globular clusters have become available. 

In this paper a new chemical evolution model is described which is designed to
account for the formation of 
both the globular clusters and the accompanying suite of field stars. The model
borrows from the classical works of galaxy formation (Eggen, Lynden-Bell, \& 
Sandage 1962, and Searle \& Zinn 1978) in that while undergoing large scale 
collapse, chemical evolution is assumed to proceed quiescently in 
protogalactic clumps which then collide as isolated units to form globular 
clusters, release stars already formed, and provide enriched and recyclable 
gas. The input to the model are the properties of two principal groups of 
globular clusters described below.

\section{Chemical Evolution and the Formation of Globular Clusters}

Possibly the most complete representative sample of the earliest local 
historical record is the system of globular clusters. For the majority of the  
clusters, we know within varying degrees of uncertainty their chemical 
abundances, motions, positions, and their ages. Unfortunately we still do not 
know how they formed. Because the evolution of individual stars is the only 
known way of enriching 
the gas from which the clusters formed, there must be a population of stars 
which formed with a distribution of heavy element abundance ending at that of 
a particular cluster. Further, because we know the heavy element distribution 
among clusters from observations, given a theory of chemical evolution we can 
calculate the abundance distribution of these accompanying stars. If as we 
shall assume below, that the clusters are formed in the shocked gas caused by 
collisions between isolated chemically evolving clumps for example, 
when the cluster is formed any remaining gas will lose energy but not angular 
momentum while the previously formed stars and accompanying dark matter will 
continue their dynamical evolution relatively unscathed. 

As discussed below 
there appear to be two groups of clusters in the Galaxy. For convenience we 
shall refer to them as the blue and red groups. The blue group is relatively 
metal poor, has low angular momentum, is spatially extended far out into the 
halo, and is generally nearly as old as the universe itself. The red group is 
relatively metal rich, shows rapid rotation at all radii, and possesses a disk 
like spatial distribution with a concentration towards the Galactic center. 
The diversity between the two groups suggests separate stages of formation and 
in our chemical evolution model each group is evolved separately. In the 
following model, the main thrust is consideration of the effects of 
globular cluster formation on chemical evolution and not the formation 
mechanism itself. For the latter subject  we refer to the more sophisticated 
modelling of Cote et al.\ (2000,2002), Beasley et al.\ (2002), and Kravtsov \& 
Gnedin (2003). Like the above cited works, ours is a bottom-up picture with 
the formation mechanism most resembling that for the metal rich clusters in  
the Beasley et al. work (i.e. collisions between gas rich clumps). For 
simplicity we assume that all clusters are formed in this way. In order to 
account for the bimodality, we argue that the collapse was anisotropic. 

The objective of this work is to reconstruct the cosmic star formation
history from local observations and in doing so to gain further insight
into how galaxies might have formed and evolved. As will be shown, this 
reconstruction is successful, comparing favorably with that deduced from 
high redshift observations (the Madau plot). This implies that our model 
provides a scenario for the chemical evolution of a representative 
sample of the universe. Further, a proposed physical framework incorporates the
results from this chemical evolution model and predicts a major new stellar
halo component, which we identify in the Milky Way with the thick disk.

\subsection{The Apparent Universality of Globular Cluster Bimodality}

Classical globular clusters like M92 belong to a group of relatively blue, 
metal poor clusters which are old and are found in the outer halo of the 
Galaxy. It was early work by Becker (1950), Baade (1958), Morgan (1959) and 
Kinman (1959) which  directed attention to the disk-like spatial distribution 
of a second group of globular clusters which are redder, more metal rich, some 
what younger, and concentrated towards the center of the Galaxy. An 
important further difference is that
whereas the metal poor group shows very little rotation about the Galactic 
center, the metal rich group shows significant rotation at all Galactocentric 
radii (Cote, 1999, Zinn 1985). A clear separation of the two groups shows in 
the frequency histogram of [Fe/H] values for 133 Milky Way globular clusters 
in Fig. 8 of Cote (1999) along with best fit Gaussians for each group.

Recent spectroscopic studies of the globular clusters in M31 by Perrett et al.\
(2002), and earlier work by Barmby et al.\ (2000) and Huchra et al.\ (1991) 
also show 
a bimodal distribution in [Fe/H] with Gaussian parameters similar to those of 
the Milky Way. In fact, the bimodality of globular cluster populations appears 
to be common in more distant luminous galaxies where it is usually 
observed as a bimodality in color distribution (e.g. Ashman \& Zepf 1998, 
Gebhardt \& Kissler-Patig 1999, Forbes et al. 2001, Kundu \& Whitmore 2001, 
Goudfrooij et al. 2003, and references therein).   

\subsection{A New Chemical Evolution Model to Accommodate 
Globular Cluster Formation}

The main ingredients of the model presented in this paper are the 
globular clusters. Unfortunately, it is still not understood in detail how 
these clusters are formed. One interesting possibility, outlined by McCrea 
(1982), suggests that clusters form in the compressed region between colliding 
streams or clumps of gas. Gunn (1980) uses a similar argument as a means to 
attain globular cluster sized Jeans masses at temperatures of a few thousand 
degrees. Some observational evidence for such a formation 
mechanism also exists (Schweitzer 1987). Given our present lack of 
understanding of how the clusters formed, we will assume the following 
phenomenological scenario. The starting point consists of a large number of 
protogalactic clumps (baryons inside dark matter halos) which, although 
spatially isolated, chemically evolve as in the simple mass loss model (a 
`quiescent' mode
of star formation). The gas in these clumps is assumed to have been initially 
enriched in a previous Population III phase. When these
clumps collide, globular clusters are assumed to form with the metal abundance
of the gas in the colliding clumps (a `burst' mode of star formation). By 
design, the mass in cluster stars is small compared to the total mass 
of all stars. After 
most collisions, quiescent star formation in each colliding clump ceases and 
the remaining gas, having lost energy but not angular momentum in the 
collision, falls towards the center of
the proto-galaxy to be recycled later. The stars previously formed quiescently
(now considered field stars) and the dark matter continue their previous 
dynamical evolution, while the newly formed clusters are not expected to have 
the same kinematics or spatial distribution as these field stars. Relatively 
few collisions could also lead to cluster formation and coalescence of the 
interacting clumps to form a satellite galaxy.

The context of the new chemical evolution model is as follows. In the well 
known mass loss model of chemical evolution described below one starts with a 
mass of gas which is slowly turned into stars while enriched gas is lost at a 
rate proportional to the star formation rate. The constant of proportionality 
determines the `effective' yield. The evolution ends when the gas is 
exhausted due to mass loss. In the new model we imagine a large number of 
`chemical evolution boxes' which individually evolve as above but may now 
collide with a neighbor. This collision is assumed to create a globular cluster
and halt the star formation in both colliding clumps. As discussed above, the 
colliding gas dissipates energy and falls to the center or as close to the 
center as the conserved angular momentum will allow. Meanwhile the stars and 
dark matter pass through and continue their previous dynamical evolution. In 
addition to the original mass loss process which can be thought of as 
`supernova driven' and hence is assumed to remove gas from further star 
formation, we now have a 
second mass loss process-the collisions-which cause the gas to lose energy and
thus be available for later recycling. Because the clusters are assumed to 
form in collisions and because the abundance distribution of each group can be 
represented by a normal distribution, in the model we make the collision rate 
normally distributed in [Fe/H] (i.e. the role of globular 
cluster formation is to effect a sudden cessation of star formation by removal
of the remaining gas from individual clumps in a Gaussian distributed manner 
in [Fe/H]).

Quantitatively, in the simple mass loss model of chemical evolution (Hartwick 
1976, see also 
Pagel 1997, p.234 and  Binney \& Merrifield 1998, p.308) the three components 
are the mass in stars, M$_{s}$, the mass in 
remaining gas, M$_{g}$, and the mass lost due to star formation, here denoted 
M$_{WHIM}$ in anticipation of later discussion. As the chemical evolution 
proceeds, the gas is gradually depleted and similtaneously enriched in heavy 
elements due to ongoing star formation, while also being lost at a rate 
proportional to the star formation rate. The 
equations governing each component for an initial mass of gas M$_{t}$ of 
metallicity Z$_{0}$, can be written as
\begin{equation}
\frac{dM_{s}}{dlog Z}={ln10}\times {Z}\times {M_{g}}/p
\end{equation}
\begin{equation}
M_{g}=M_{t}\times{exp(-(Z-Z_{0})(1+c)/p)}
\end{equation}
\begin{equation}
M_{WHIM}={c}\times{M_{s}}
\end{equation}
\noindent where M$_{s}$(Z$_{0}$)=0 and M$_{g}$(Z$_{0}$)= M$_{t}$, p is the 
yield of heavy elements and c is the ratio of mass loss rate to
star formation rate and also determines the `effective' yield (i.e. $p_{eff}=p/
(1+c)$).

In our new model, it is assumed that the collisions between clumps which  
halt the chemical evolution are distributed normally with mean log Z$_{c}$ and
standard deviation $\sigma$. As the collisions occur, the mass in surviving 
clumps diminishes. This surviving clump mass with metallicity 
greater than log Z is M$_{t}\times$~f~($>$~log Z) where the initial mass of 
gas in all clumps with metallicity Z$_{0}$ is M$_{t}$ and
\begin{equation}
f(>~log Z)\simeq\frac{1}{\sqrt{2\pi}\sigma}\int_{log Z}^{\infty} exp\left(-
\frac
{(log Z^{\prime}-log Z_{c})^{2}}{2\sigma^{2}}\right)~d log Z^{\prime}
\end{equation}
\noindent which is valid when Z$_{0}$ $\ll$ Z$_{c}$.
Note that for computing purposes the right hand side of (4) is one 
half of the complementary error function i.e.
$f~(> log Z)=0.5\times erfc~\left(\frac{(log Z-log Z_{c})}{\sqrt{2}\sigma}
\right)$. By replacing M$_{t}$ with M$_{t}\times$~f, equations (1) and (2) can
be written as 
\begin{equation}
\frac{dM_{s}}{dlog Z}={ln10}\times {Z}\times {M_{g}}/p
\end{equation}
\begin{equation}
M_{g}={M_{t}}\times{0.5}\times{erfc~\left(\frac{(log Z-log Z_{c})}{
\sqrt{2}\sigma}\right)}\times{exp(-(Z-Z_{0})(1+c)/p)}
\end{equation}
\noindent Note that equations (1) \& (3) remain the same. (Equation (1) remains
the same because the metallicity of the gas lost by collisions is the same as 
that in the clumps at the time of the collision). Because we now have an 
additional mass loss component, we define a new variable M$_{ml}$, which 
represents the gas lost when the clumps collide. Whereas M$_{WHIM}$ is 
considered here to be lost to the `warm hot intergalactic medium', the new 
M$_{ml}$ component, having lost energy in the collision, is assumed to 
fall to the center of the proto-galaxy to be recycled. Figuratively, an
individual classical chemical evolution `box' is totally purged by having 
gas drop out of the bottom on collision as well as having been blown 
continuously out of the top as a result of previous ongoing star formation!
The equations governing the evolution of these two mass loss components are
\begin{equation}
M_{WHIM}={c}\times{M_{s}}
\end{equation}
\noindent and
\begin{equation}
\frac{dM_{ml}}{dlog Z}={M_{t}}\times{exp(-(Z-Z_{0})(1+c)/p)}\times{
\frac{1}{\sqrt{2\pi}\sigma}}\times{exp\left(-\frac{(log Z-log Z_{c})^
{2}}{2\sigma^{2}}\right)}
\end{equation}
We limit the number of parameters by letting Z$_{c}$ define the 
effective yield, i.e.
\begin{equation}
c=(p/Z_{c}-1)
\end{equation}

Equation (5) represents the MDF of all stars formed (including those in
globular clusters). We have not attempted to model globular cluster formation.
It is assumed that the clusters are formed in bursts during collisions between 
clumps. We do attempt to make the chemical evolution consistent by ensuring 
that stars are formed which, if assembled into a cluster system, would have 
the requisite mass and (by design) the appropriate Gaussian abundance 
distribution. To this end we express the 
amplitude of the cluster MDF as the product of the $total$ baryonic mass 
(M$_{t}$) (e.g. McLaughlin, 1999) and an efficiency factor $\eta$ so that  
\begin{equation}
\frac{dM_{GC}}{dlogZ}=\eta M_{t}\frac{1}{\sqrt{2\pi}\sigma}
{exp\left(-\frac{(log Z-log Z_{c})^{2}}{2\sigma^{2}}\right)},
\end{equation}
\noindent where $\eta=0.00068$. This value of $\eta$ was chosen in order
to ensure the assembly of $\sim100$ metal poor (blue) clusters in the model
to be discussed later. For the parameters adopted in the model to follow, all 
but the most metal rich 3\% of the the blue clusters and 99.8\% of the same 
number of metal rich (red) clusters can be accommodated before the gas is 
exhausted. Since the M$_{ml}$ component eventually gets turned into stars, the
ratio of the integrated mass in clusters to that in stars 
(i.e. (M$_{GC}$/(M$_{s}$+M$_{ml}-$M$_{GC})$) is $\sim0.001$ which is within a 
factor of two of observational determinations of the same quantity (e.g. 
McLaughlin, 1999).

We reiterate that in the context of this model the formation of globular 
clusters only determines the extent of the chemical evolution within 
individual clumps. It is the stars produced that enriched the gas from which 
the clusters formed and the enriched gas lost both due to star formation 
and collisions which are the dominant end products in this model and not 
the individual clusters.

Equations (5)-(10) along with boundary conditions M$_{s}(Z_{0})=0$, M$_{g}
(Z_{0})$=M$_{t},$ M$_{WHIM}(Z_{0})=0,$ M$_{ml}(Z_{0})=0$ and M$_{GC}
(Z_{0})=0$ and parameters $Z_{c}$, $Z_{0}$, $\sigma$ and $\eta$ define the new 
chemical evolution model. Figure 1 shows the evolution of 
the independent quantities for an illustrative set of 
parameters. The parameters were chosen to provide observable separation of
the four components, but it should be emphasized that the model was motivated
such that log~Z$_{c}$ and $\sigma$ are determined by a particular 
$globular$ $cluster$ abundance distribution. As a result of the loss of gas  
due to collisions, the peak in the accompanying 
$stellar$ metal abundance distribution is found to be $\sim0.25$ dex 
$lower$ than the corresponding globular cluster peak (log~Z$_{c}$) for values 
of $\sigma$ of order 0.2-0.3.

\subsection{Colliding Boxes and the Stellar Halo of M31}

Durrell et al.\ (2001) have probed the halo of M31 at a distance of $\sim$20 
kpc
from the center and have determined the metallicity distribution function 
(MDF) photometrically. 
This study provides an excellent `test' sample for the new model, since
all stars are essentially at the same distance from the Milky Way and the
photometry is sufficiently deep that there should be no serious observational
selection effects. Their results are reproduced as the histogram in Figure 2,
where for consistency their metallicity scale [m/H]  has been converted to 
[Fe/H] by subtracting 0.3 dex (as described in their paper). In a recent 
spectroscopic study of a large sample of globular clusters in M31, Perrett et 
al.\ (2002) confirm earlier work 
that the abundance distribution is bimodal. These authors fit Gaussians with 
[Fe/H]$=-1.44~(\sigma^{2}=0.22)$ and [Fe/H]$=-0.5~(\sigma^{2}=0.13)$ to their
observations. The solid
line in Figure 2 shows the MDF predicted by the new model assuming the sum of 
contributions from a log$Z_{c}/Z_{\odot}=-0.55~(\sigma=0.2)$ component and 
from a log$Z_{c}/Z_{\odot}=-1.5~(\sigma=0.3)$ component in the initial ratio 
of 2.4:1 and normalized to the same area as the observations. The fit is 
reasonable while  possessing
the feature of unifying the clusters and stars of each population. Note
the offset in the peak of the stellar distribution from the assumed value of
log$Z_{c}$ of the metal-rich component.

The simple phenomenological picture outlined above now allows us to model early
galaxy evolution based on the near universal appearance of two groups of 
globular clusters. However, in order to use the model to determine galaxy 
evolution as a function of time and/or redshift and to calculate star 
formation rates, we need a relationship between metal abundance and age.

\section{The Age-Metallicity Relation for the Globular Clusters}

In order to determine the age-metallicity relation (AMR) for globular
clusters, we refer to the recent papers by VandenBerg (2000) and Salaris \& 
Weiss (2002). The two studies were
combined in the following manner. A zero point of 13.5 Gyr was set for the 
cluster M92 based on the work of Vandenberg et al.\ (2002). This required 
reducing all the ages in the VandenBerg (2000) sample by 0.5 Gyr. A comparison 
with clusters in common with the Salaris \& Weiss sample was then 
made  and 0.6 Gyr was added to each cluster in
this work. Wherever the two studies had clusters in common the VandenBerg
ages were used. It will be seen later that imposing this zero point on cluster
ages allows only a very narrow range of consistent values of the Hubble 
constant. These data are
shown plotted in Figure 3 against new determinations of [Fe/H] by Kraft \& 
Ivans (2003a, 2003b). The importance of this new globular cluster [Fe/H] scale
is that the iron abundances were determined from FeII lines rather than FeI 
and arguably should be more reliable. The uncertainties in the cluster ages 
are large, but the scatter appears to increase
at [Fe/H]$\geq-1.6$. In order to account for this apparent increase in scatter
in the simplest way, a bifurcated relation was adopted.
The form of the relations assumed are 
\begin{equation}
          Z/Z_{\odot}=\phi(x/x_{s})^{-{\alpha}}exp(-(x/x_{s})^{\beta})
\end{equation}
where $x=10^{t_{9}}$ and $t_{9}$ is the lookback time in billions of years.
The red cluster relation which extends to the highest [Fe/H] has 
parameters 
\begin {eqnarray}
\phi=0.065~~~~~x_{s}=10^{13.08}~~~~~\alpha=0.15~~~~~\beta=1.0  
\end{eqnarray}
The blue cluster relation follows the above for [Fe/H] $<-2.19$, but for
larger [Fe/H] the parameters are:
\begin{eqnarray}
\phi=0.0195~~~~x_{s}=10^{13.40}~~~~~\alpha=0.15~~~~~\beta=2.0
\end{eqnarray}
These relations are shown superimposed on the data in Figure 3.

Later an age-metallicity relation will be required for modelling the bulge 
star formation rate. It is constrained by the observations of Zoccali et al.\ 
(2003) showing there are no bulge stars with ages $<10^{10}$ yrs. The relation 
adopted is shown as the dashed line in Figure 3. Its parameters (valid for 
[Fe/H]$>-1.6$) are:
\begin {eqnarray}
\phi=0.065~~~~~x_{s}=10^{13.08}~~~~~\alpha=0.60~~~~~\beta=1.0  
\end{eqnarray}

\subsection{Calculating the Star Formation Rate}

Equation (5) allows calculation of d$M_{s}$/dlog Z, while equation 
(11) allows calculation of dlog Z/dt. The product of these two
quantities then yields the star formation rate d$M_{s}$/dt. From (11)
\begin{equation}
\frac{dlog Z}{dt_{9}}=-\alpha-\beta(x/x_{s})^{\beta}
\end{equation}
Since the derivation of equations (5)-(10) implicitly assumed instantaneous
recycling and $M_{s}$ represents the long lived stars and remnants, the actual 
SFR will be 
\begin{equation}
\frac{dS}{dt}=\frac{1}{(1-R)}\frac{dM_{s}}{dlogZ}\left(-\frac{dlogZ}{dt_{9}}
\right),
\end{equation}
\noindent where S(t) is the mass of $all$ stars formed up to time t, R is the 
return fraction (Pagel 1997, p.213) and t is the $evolution$ time in billions
of years. We have assumed ($1-$R)$=0.7$, appropriate for a Salpeter IMF (Madau
et
al.\ 1998). Note that the units of (16) are M$_{\odot}$/yr, if the unit of M$_
{t}$ in equations (5)-(10) is $10^{9}$ M$_{\odot}$.

\subsection{Specifying the Cosmology: The Lookback Time-Redshift Relation}

In order to relate the lookback time and hence [Fe/H] via equation (11) to
redshift z, we need to specify a cosmological model. The model chosen has
$\Omega_{m}=0.3$ and $\Omega_{\Lambda}=(1-\Omega_{m})$. The general relation
between 
time and redshift can be found in Peebles (1993, equation 13.20). 
Letting the Hubble constant be H$_{0}=100h$ km~s$^{-1}$Mpc$^{-1}$, 
for the above parameters the age of the universe is 9.426/h and 
\begin{equation}
z=[0.6547sinh(1.201(1-\frac{t_{9}h}{9.426}))]^{-\frac{2}{3}}-1
\end{equation}
\begin{equation}
t_{9}=\frac{9.426}{h}[1-0.8265~ln(1.5275(1+z)^{-\frac{3}{2}}+(2.333(1+z)^
{-3}+1)^{\frac{1}{2}})]
\end{equation}
\noindent The unit of the age and lookback time, $t_{9}$, is $10^{9}$ years.

\section{The Galaxy Formation Scenario}

The goal of this investigation is to re-construct the Madau plot from local
observations. This is done by using the above model to chemically evolve
a representative co-moving volume of the universe to form what will be a 
representative or composite galaxy. It may be more than fortuitous that Baldry
et al.\ (2002)
showed that the averaged spectra of 2DF galaxies as a function of redshift 
resemble what would be a present day Sb-Sbc galaxy (i.e. similar to M31 and the
Milky Way, galaxies on which our model has been `calibrated').

\subsection{The Spheroidal Component}

The starting point is to define a representative co-moving volume of the 
universe which we take to be 100 Mpc$^{3}$. The volume assumed is arbitrary 
and has no 
effect on the final result. However, we note that $\phi^{*}$ of a generic 
galaxy luminosty function is $\sim0.01 Mpc^{-3}$, implying a volume per $L^{*}$
galaxy of $100 Mpc^{3}$. Given $\Omega_{b}h^{2}=0.0224$ from WMAP (Spergel et 
al.\ 2003), this volume encloses a baryonic mass of 6.22x10$^{11}$ M$_{\odot}$
. While as yet we have no way of knowing the mass in remnants left from a 
Population III phase, it is assumed
that within this volume there is at least $1.2\times10^{11}$ M$_{\odot}
$ of gas which has been enriched during this earlier phase to [Fe/H]$=-4.0$. 
These numbers make up the M$_{t}=$(M$_{t}$(blue)+M$_{t}$(red)) and Z$_{0}$ 
of our model. To form what 
will be the main spheroidal component, we assign the parameters shown in 
Table 1. As can be seen from the fourth column, most of the parameters are
observationally constrained. The cluster parameters, log $Z_{c}/Z_{\odot}$ and 
$\sigma$, come from Cote's discussion, and the ratio of M$_{t}$(blue) to M$_{t}
$(red) is the same value deduced from the discussion in \S2.3 of the 
M31 halo from observations by Durrell et al. The adopted value of the 
yield, p$=0.013$ (log(p/Z$_{\odot})=-0.11$), is determined from a study of 
Galactic bulge stars (Zoccali et al.\ 2003) which will be discussed in more 
detail 
in the next section. It should be noted that observational evidence suggests
that field halo, cluster stars, thick disk, and bulge stars all exhibit an
enhanced oxygen to iron abundance ratio (e.g. Prochaska et al.\ 2000). This 
provides the justification for modelling the evolution of the iron abundance 
under the instantaneous recycling approximation. 

Using the parameters given above and equation (16) to calculate the SFR, the
results are shown in Figure 4. The final products of the chemical evolution
phase are for the blue component M$_{s}=0.719$, M$_{ml}=13.1$, and M$_{WHIM}=
21.2$ and for the red component M$_{s}=19.6$, M$_{ml}=31.7$, and M$_{WHIM}=
33.6$, all in units of $10^{9}$ M$_{\odot}$. We note from Figure 4 that whereas
the SFR of the blue component has ended by t$_{9}\sim13$ Gyr, the model
predicts that the red spheroidal
component is still forming stars at t$_{9}\sim6-7$, remarkably consistent with
recent observations of M31 halo stars by Brown et 
al.\ (2003). In a later section these results
will be discussed within the context of a `toy' model for galaxy 
formation. Anticipating the main results of this discussion, both M$_{s,blue}$
and M$_{s,red}$ will populate the outer halo, M$_{ml,blue}$ will become the 
bulge, M$_{ml,red}$ will form the disk, while both components of M$_{WHIM}$ are
assumed to be lost from the galaxy.

\begin{deluxetable}{cccc}
\footnotesize
\tablenum{1}
\tablecolumns{4}
\tablecaption{Model Parameters}
\tablehead{
\colhead{Model/equations} & 
\colhead{Parameter} &
\colhead{Value} &
\colhead{Source}  
}
\startdata
5-10 & log(Z$_{0}$/Z$_{\odot}$)$_{blue,red}$ & $-4.0\pm1.0$  & assumed \nl
$\prime\prime$ & M$_{t}=$(M$_{t,blue}+$M$_{t,red}$) & $1.2\pm0.24\times10^{11}$
M$_{\odot}$ & High z observations in Fig. 5 \nl
$\prime\prime$ & M$_{t,red}$/M$_{t,blue}$ & $2.43\pm0.10$ & Durrell et al.\ 
(2001) \& \S2.3 \nl
$\prime\prime$ & log(Z$_{c}$/Z$_{\odot}$)$_{blue}$ & $-1.6\pm0.1$ & Cote (1999)
\nl
$\prime\prime$ & $\sigma_{blue}$ & $0.3\pm0.1$ & Cote (1999) \nl
$\prime\prime$ & log(p/Z$_{\odot}$)$_{blue}$ & $-0.11\pm0.03$ & Zocalli et 
al.\ (2003) \nl
$\prime\prime$ & M$_{WHIM,blue}$ & $2.1\pm0.42\times10^{10}$M$_{\odot}$ & from
 model \nl
$\prime\prime$ & M$_{s,blue}$ & $0.072\pm0,014\times10^{10}$M$_{\odot}$ & from
 model \nl
$\prime\prime$ & log(Z$_{c}$/Z$_{\odot}$)$_{red}$ & $-0.55\pm0.1$ & Cote (1999)
 \nl
$\prime\prime$ & $\sigma_{red}$ & $0.3\pm0.1$ & Cote (1999) \nl
$\prime\prime$ & log(p/Z$_{\odot}$)$_{red}$ & $-0.11\pm0.03$ & Zocalli et al.\
 (2003) \nl
$\prime\prime$ & M$_{WHIM,red}$ & $3.4\pm0.67\times10^{10}$M$_{\odot}$ & from 
model \nl
$\prime\prime$ & M$_{s,red}$ & $1.96\pm0.39\times10^{10}$M$_{\odot}$ & from
 model \nl
1-3 & M$_{t,bulge}$ & $1.3\pm0.26\times10^{10}$M$_{\odot}$ & M$_{ml,blue}$ 
from model \nl
$\prime\prime$ & c$_{bulge}$ & 0 & Zocalli et al.\ (2003) \nl
$\prime\prime$ & log(p/Z$_{\odot}$)$_{bulge}$ & $-0.11\pm0.03$ & Zocalli et 
al.\ (2003) \nl
19 & M$_{t,disk}$ & $3.2\pm0.62\times10^{10}$M$_{\odot}$ & M$_{ml,red}$ and 
\S4.3 \nl
17-18 & Cosmology ($\Omega_{m},\Omega_{\Lambda}$, h) & 0.3, 0.7, 0.66 & assumed
\nl
& Z$_{\odot}$ & 0.017 & \nl
\enddata

\end{deluxetable}

\subsection{The Bulge Component}

Although the bulge is usually considered to be part of the spheroidal 
component of a galaxy, it will be discussed separately here. Recently Zoccali 
et al.\ (2003) 
have determined the metallicity distribution of a sample of Galactic bulge 
stars and find that it resembles the distribution resulting from closed box
chemical evolution (equation (1) \& (2) with c=0) with log Z$_{0}$/Z$_{\odot}
\sim-1.6$
and log p/Z$_{\odot}\sim-0.11$. (The mean metallicity of M$_{ml,blue}$ is 
predicted to be similar to this value of $Z_{0}$ while the the yield implied 
is the same as that assumed in the model). Further, the above 
authors also 
point out that there are no stars with ages $<10^{10}$ yrs in their bulge 
sample. In order to calculate the SFR associated with this metallicity 
distribution, the metal-rich component of the adopted age metallicity relation
(equations (11) \& (14)) and equation (16) were
evaluated with the closed box assumption (equation (1) \& (2) with c=0) for 
the dM$_{s}$/dlogZ derivative. The results are shown in Figure 4. The peak in 
the SFR occurs at t$_{9}\sim11.3$ and becomes negligble at t$_{9}<10$, 
in accord with the Zoccali et al.\ results.

\subsection{The Disk Component}

Just, Fuchs \& Wielen (1996) have developed a method for determining the 
$local$ disk SFR by modelling the vertical structure of the Galactic disk. 
Their best model shows a sharp initial maximum followed by a slow decline 
(Just 2002). In addition, recent infrared studies of the populations in the 
central parts of the Galaxy (e.g. van Loon et al.\ 2003), show a very  
active star formation history. These authors find that the inner Galaxy is 
dominated by an old population ($\geq7\times10^{9}$ yrs) plus an 
intermediate age population ($2\times10^{8}-7\times10^{9}$yrs) and even 
younger 
stars. These results have prompted the construction of a two component model to
represent the disk SFR. The first is a constant SFR of 1 M$_{
\odot}$/yr for $10^{10}$ yrs with a Gaussian tail with $\sigma=0.6$. For t$_{9}
\leq10$, the second component is given by
\begin{equation}
\frac{dS}{dt}=1.5\times10^{0.45z(t_{9})},
\end{equation}
\noindent while for t$_{9}>10$ a Gaussian tail with $\sigma_{t_{9}}=0.6$ was
assumed. The sum of these two components is shown in Figure 4. Integrating 
the rate over time and multiplying the result by the assumed value of ($1-$R) 
$=0.7$ gives a model equivalent M$_{s}$ of 3.2x10$^{10}$
M$_{\odot}$, which is identical with M$_{ml,red}$ (Table 1) and hence 
consistent with our model. We estimate that the
ratio of present SFR to average SFR for this disk component is $\sim0.6$, 
which is well within the range expected for an Sb-Sbc galaxy, according to 
Table 8.2 of Binney \& Merrifield (1998) which is based on data from
Kennicutt et al.\ (1994).

We note from Figure 4 that there is a relatively large overlap in SFR between 
the disk and the red spheroidal component. Star formation associated with the 
latter component goes on until t$_{9}\sim6-7$ Gyr ($\S4.1$), implying that the
last protogalactic clumps to arrive must have originated at a great distance 
from the galactic center. This late arrival would allow 
time for dynamical friction to cause the still quiescently evolving clumps to 
spiral in to the center of the forming galaxy before colliding to make metal 
rich globular clusters and to provide gas for the observed ongoing star 
formation towards the inner galaxy. To observers looking towards the Galactic 
center, this relatively young, relatively metal rich, relatively high angular 
momentum `disk' component could easily appear to be associated with the 
bulge. One could then understand why the inner metal rich 
clusters (R$_{g}<4$ kpc) appear to be associated with the bulge even though 
they possess significant rotation while the outer red clusters exhibit disk 
characteristics (Cote 1999, Forbes et al.\ 2001). 

Further, from equation (10), the ratio of mass in red
clusters to blue clusters should be equal to the ratio M$_{t,red}$/M$_{t,blue}
$ which from Table 1 is $\sim2.4$. As the observed red to blue cluster ratio is
closer to unity, we must assume that either many of the red clusters which do 
form are disrupted or that the efficiency of red cluster formation is  
lower due perhaps to tidal effects and to the presence of a substantial 
existing disk 
component. In a later discussion, we will attribute the surviving red clusters
to a thick disk component. Also part of the thick disk are the stars from 
disrupted clusters and the earliest formed disk stars resulting from the 
compression of in situ
disk gas from late collisions with still intact protogalactic clumps 
(assumed to be responsible for the initial burst in the disk SFR at 10 Gyr).  

\subsection{The Warm Hot Intergalactic Medium Contribution}

As the model is currently defined, M$_{WHIM}$ represents the mass loss 
associated with star formation. The mean metallicity of each component 
(M$_{WHIM,blue}$ and M$_{WHIM,red}$) will be
the same as the respective stellar components. Assuming that
these WHIM components are mixed, we find a total mass of $5.5\pm0.80\times10^
{10}$M$_{\odot}$ with a mean [Fe/H] of $\sim-1$. As little is presently known 
concerning the hot gas surrounding an individual galaxy, we refer to the 
situation in small groups of galaxies. Even here, the metal abundance of hot 
gas in
small groups of galaxies is poorly known, but the lowest determined values are
consistent with the above mean (Mulchaey 2000). The ratio of the mass in hot
gas (the WHIM) to stellar mass is $\sim0.9\pm0.2$, a value similar to what may
be only a lower limit deduced from x-ray observations of hot gas in $groups$ 
of galaxies (Mulchaey 2000). Another statistic discussed by Mulchaey, is the 
ratio of baryonic mass to total mass which can be evaluated as $\rho(M_{t})/
(\Omega_{m}\rho_{crit})$. Our model gives $\sim0.04\pm0.01$,  
which is also the lower end of the distribution found for small groups. 
Finally, we determine $\Omega_{WHIM}=0.0048\pm0.0007$ which compares favorably 
with the value of $\gtrsim0.0046~h^{-1}_{0.66}$ from Tripp et al.\ (2000).

\subsection{The History of the Star Formation Rate Density from Local 
Observations}

A Madau plot is now easily constructed by adding the SFRs shown in Figure 4
together, dividing by the co-moving volume, taking the logarithm, and plotting
the result against the corresponding redshift z(t$_{9}$). This is shown in
Figure 5 along with the data obtained from the high redshift observations. The
sources for the observational data, which have been adjusted to our assumed
cosmological model and corrected for absorption, are given in the figure 
caption. Although the model is made up of only four components, there are many 
parameters
involved, and the model is certainly not unique. In theory, there are only
two free parameters: M$_{t}$ and log Z$_{0}$. Varying log Z$_{0}$ by $\pm1.0$ 
does not change the plot, while a $20\%$ change in M$_{t}$ results in $\sim20
\%$ change in overall normalization. In addition, the figure is insensitive to 
changes of $\pm0.1$ in each of log Z$_{c,blue}$, log Z$_{c,red}$, $\sigma_
{blue}$, 
$\sigma_{red}$, and the ratio M$_{t,red}$/M$_{blue}$. Varying the yield, p, by
$\pm0.03$ results in $\sim\mp23\%$ changes in overall normalization. The shape
of Figure 5 is dependent on the derivatives of the 
age metallicity relations adopted (equation (15)), and they become very large 
and rapidly 
changing at low [Fe/H]. Finally, the slope of the high redshift tail of Figure 
5 is dependent on the adopted value of h for a given globular cluster age zero
point. The masses of the various components obtained from our successful 
reconstruction of the Madau plot should be relatively robust and allow 
additional consistency checks on the plausibility of the model. An uncertainty
of $\pm20\%$ in the value of M$_{t}$ is assumed and is propagated throughout.

From the model we find $\Omega_{\star}=0.0056\pm0.0007$, which can be compared
with the Fukugita, Hogan \& Peebles (1998,FHP) value of $\Omega_{\star}=0.0037^
{+0.0019}_{-0.0013}$h$^{-1}_{0.66}$. Although the error bars overlap, our value
is $\sim50\%$ higher. Further, the model shows that M$_{disk}$/M$_{spheroid}=
0.9\pm0.2$ 
compared to $\sim0.3$ from FHP, $\sim1$ from Schechter \& Dressler (1987), and 
$\sim1.3\pm0.2$ from Benson et al.\ ( 2002). Finally, the gas from the 
Population III phase that did not go into making the galaxy is considered to 
be a diffuse 
component and contributes $\Omega_{diffuse}=0.044\pm0.002$. (This assumes that
the mass in dark condensed remnants is negligible). Here it is considered 
separate from the WHIM and tentatively identified with the gas responsible for
producing the local Lyman $\alpha$ forest which apparently could significantly
exceed 32\% of all local baryons (Stocke et al.\ 2003).

Two additional consistency checks on our model are now considered. The first 
is concerned with the origin of the damped Lyman $\alpha$ systems (DLA's) for 
which any complete model of chemical evolution applied to the entire universe 
should account. Figure 6 shows the assumed age-metallicity relations 
(converted to metallicity-redshift relations) on which are superimposed a 
selection of observations of DLA's from Kulkarni \& Fall (2002) and Prochaska 
et al.\ (2003). Recall that the two relations apply to each of the metal poor 
clusters, and metal rich clusters.  In order to make a
robust statistical estimate of the probability that any particular component 
is intersected by a random line of sight, one needs to know the distribution
in area as well as the number and size of the protogalactic clumps, parameters 
which are not specified by the model. At most, we can calculate the amount of 
gas remaining in each component as a function of [Fe/H]. The range over which 
this amount is $>$1\% of the original amounts of 3.5, and 8.5$\times10^{10}$M
$_{\odot}$ is indicated by the thickened parts of the lines in 
Figure 6. As it stands the model does not predict either an age-metallicity 
relation nor a metallicity distribution function for the disk component, but 
only the mass of gas that is available, $3.2\times10^{10}$M$_{\odot}$. This 
comparison with the observations of DLA's is necessarily crude, but
there appear to be no major inconsistencies. 

Secondly, we note that both the large reservoir of low angular momentum gas 
released after the first collapse and a conduit to the center of the galaxy 
(along the rotation axis) are ingredients required for fuelling and possibly 
creating a black hole. Thus, the model naturally provides an environment 
conducive to the establishment of a relationship between the mass of a 
central black hole and of the surrounding bulge stars. If M${\bullet}$/M$_
{bulge}=0.002$ (Magorrian et al.\ 1998), then the mass of the black hole in the
model is M${\bullet}=2.6\pm0.5\times10^{7}$M$_{\odot}$, comparable to the 
black hole mass in the nucleus of M31. When divided by the assumed 
comoving volume one finds $\rho_{\bullet}=2.6\pm0.5\times10^{5}$M$_{\odot}$
/Mpc$^{3}$ for the local black hole density which is in excellent agreement 
with the recent 
determination of this quantity by Yu \& Tremaine (2002). Further, the maximum 
SFR of the bulge component occurs at redshift z$\sim2.3$, which is close 
to the redshift at which the space density of luminous quasars is a maximum.

\subsection{Colliding Boxes and Galaxy Formation}

The main quantitative results of this investigation are contained in Figure 5 
and Table 3.
What follows is a proposed $physical$ framework which could accommodate the 
chemical evolution scenario described above.  Basically, we envision the 
anisotropic collapse of a number of initially isolated, quiescently evolving, 
protogalactic clumps. The initial configuration is then either a triaxial halo
(CDM) or a fragmented sheet or flattened filament (WDM). Evidence which 
supports
these initial conditions is the triaxial spatial distribution of the outlying
satellites of the Galaxy (Hartwick 2000) along with a similar alignment of the
space distribution and 
space motions of the blue globular clusters (Hartwick 2002). During the first 
collapse, which is $transverse$ to the long (rotation) axis, the 
low angular momentum clumps will collide to form the blue globular clusters. 
In turn, the gas released in the collisions will fall towards the center to 
form the bulge. Support for a low angular momentum halo-bulge 
connection can be found in the review of 
Wyse et al.\ (1997), who show (their Figure 7) that the distribution of the 
specific angular momentum of the (classical) halo and bulge are similar and 
unlike the disk/thick disk distribution.
The higher angular momentum clumps will generally not collide initially,  
but instead will fall towards the center along the long axis while still 
forming stars and enriching the gas within them. These clumps will
eventually meet at what will become the plane of the disk. Red clusters 
then form in these collisions, and the gas released settles down to form the 
disk. The stars which had previously formed continue their dynamical evolution
back out into the halo. The MDF of this group of 
stars should be similar to the dashed curve in Figure 2, with a peak at [Fe/H]$
\sim-0.8$ and with an angular momentum distribution like the thin disk. 
Note that there is a metal poor tail associated with this MDF. {\it The 
existence of a relatively high angular momentum (disk-like), relatively metal 
rich, and  relatively massive component of the Galactic halo is an inescapable
(and falsifiable) prediction of this model.} 

In \S4.3 we noted that according to our model there should be $\sim2$ times 
more red clusters than blue being formed. However, given the fact 
that the SFR of the disk and red component overlap (Figure 4), the later 
infalling clumps are more likely to collide with existing disk gas rather than 
other clumps or be tidally disrupted. It is assumed that not only will many 
fewer red clusters be 
formed, but that the compression of this gas from infalling clumps will form 
stars 
whose kinematics can be expected to resemble the red clusters which survive  
in numbers comparable to blue clusters. (In equation (10) $\eta$ 
is effectively reduced by $\sim1/2$). A thick disk component is thus
created. The peak in the disk SFR at 10 Gyr in Figure 4 is identified with 
this process, and the mass in the thick disk was estimated by integrating over
this peak to t$_{9}\sim9.1$. The result is M$_{TD}=1.04\times10^{10}$M$_
{\odot}$. 

A further
consequence of the overlap in SFR referred to above is that there will also be 
an overlap both in age and metal abundance among metal poor thick disk stars 
and metal rich red halo stars. An important difference is that the thick disk 
as defined above will $not$ possess a metal poor tail unlike the red halo 
stars. In addition to the metal rich clusters, the thick disk could also be 
made up of the first stars to form 
from the $gas$ of the M$_{ml,red}$ component with mean $\left<[Fe/H]\right>
\sim-0.7$, $\sigma\sim0.3$ (i.e. there should be relatively few stars with 
[Fe/H]$\leq-1$) so that at a minimum the MDF might resemble that of the metal 
rich clusters ($\left<[Fe/H]\right>\sim-0.55$, $\sigma\sim0.3$). The disk 
component is not chemically (nor dynamically) evolved here so it is not 
possible to predict just how kinematically and chemically distinct will be the
stellar component formed in the spike of the disk SFR at 10 Gyr in Figure 4  
which we consider to be part of the thick disk (but see the discussion 
below for a possible Milky Way counterpart).

Except for one possible caveat (dynamical evolutionary effects such as 
dynamical friction for example) the vertical scale height of the blue halo 
component (including the blue globular clusters) should be comparable to the 
red halo component so that the metal poor blue globular clusters and the metal
rich red halo field stars 
will occupy the galaxy halo contemporaneously. This result could provide a 
natural explanation for the observations of Forte et al.\ (1981) who showed 
that the integrated light from halo stars was redder than that of the globular
clusters at the same projected radius in four galaxies in the Virgo cluster. 
Subsequently, similar results have been found in other galaxies including the 
Milky Way and M31. The effect is attributed to a difference in metal abundance
between the two populations (e.g. Harris, 1991) in agreement with the 
prediction. Below we will argue that the above caveat applies to the Milky Way
with the result that the red component is more flattened than the blue.

Table 2 summarizes the relative sizes of the various components of our 
framework 
galaxy. More sophisticated modelling is required to turn the mass ratios in 
Table 2 into local density ratios. It should also be emphasized that this is
a composite model based on observations of the Galaxy and M31 although it 
likely more closely resembles M31. While we 
expect all of its components to be present in both galaxies, the actual ratios 
will undoubtedly be different. 
\begin{deluxetable}{cllc}
\tablenum{2}
\tablecolumns{4}
\tablecaption{A Framework Model}
\tablehead{
\colhead{} &
\colhead{Component} & 
\colhead{M/M$_{disk}$} &
\colhead{}   
}
\startdata
 & Disk & 1.0 & \nl
 & Thick Disk & 0.33 & \nl
 & Surviving Red Globular Clusters & 0.000751 & \nl
 & Bulge & 0.42 & \nl
 & Red Stellar Halo & 0.63 &  \nl
 & Blue Stellar Halo & 0.023 & \nl
 & Blue Globular Clusters & 0.000751 & \nl
\enddata
\end{deluxetable}

We now consider whether the predicted red halo component and the thick disk as
defined by the framework model actually do have counterparts in the Milky Way.
A high latitude deep CCD survey has recently been discussed by Karaali 
et al.\ (2003) (see also Chen et al.\ 2001). By using ultraviolet excess as a 
metallicity indicator they 
show (their Fig. 11) the distribution of stars as a function of metallicity in 
progressively fainter magnitude bins. The diagrams show 3 peaks at [Fe/H]
$\sim$-0.06, -0.83, and -1.59 which the authors attribute to thin disk, thick 
disk, and halo respectively. Comparing these diagrams with Fig. 4 of Beers 
(1999) which shows the (admittedly biased) MDF of metal poor star candidates,  
we see very good correspondence with the two metal poorest distributions.These
MDF's are just what our two component halo model predicts, although the red 
halo component apparently manifests itself as the thick disk component of Fig.
11 above. This identification would allow a natural explanation for the `metal 
weak thick disk' stars (e.g. Morrison et al.\ 1990, Beers et al.\ 2002) and 
the flattened component of the `classical' halo tracers as represented by 
the RR Lyrae stars (Hartwick, 1987) and blue horizontal branch stars (Kinman 
et al.\ 1994) as the expected metal poor tail of the red halo component. The 
observed flattening is most likely a result of dynamical evolution 
and hence is not predicted by our framework model which is chemical 
evolution based. Binney \& May (1986) have shown that a thick disk 
structure is expected if the angular momentum of the stars formed in the 
infalling proto-disk material is not parallel to a principal axis of the 
early spheroid. For the Milky Way, some evidence for such a misalignment 
exists (Hartwick, 2000, 2002). Whatever the 
mechanism, the apparent flattening along with the predicted relatively high 
angular momentum of the red component, when observed against a background of 
the blue (classical low angular momentum) stellar halo component, should show 
a gradient in circular velocity with respect to height above the plane which 
is qualitatively consistent with that seen in Fig. 4 of Chiba \& Beers (2000).
Carney (2001) finds a similar result (his Fig. 3) and concludes that there are
at least two populations of metal poor stars one of which is flattened and 
rotating relatively rapidly.

On the basis of the above discussion, we expect the bulk of the stars of the 
red halo component to occupy a flattened distribution. However, it would not be
surprising if some of the clumps, while being dragged down, became tidally 
disrupted, resulting in some metal rich stars being 
found at large distances from the Galactic plane. Ratnatunga and Freeman (1989)
showed that at $\sim5$ kpc above the plane the number of K giants with 
[Fe/H]$\geq-1$ is 12 out of 32. Recently, Morrison et al. (2003) presented 
preliminary results of a more distant survey for halo stars and find 5 out of 
30 stars with [Fe/H]$\geq-1$ at distances of order 30 kpc.

As emphasized earlier, the $model$ thick disk does not have a metal poor tail
and hence is different from the thick disk identified above. In our Galaxy, we
identify its `backbone' with the metal rich globular clusters and with the 
observed stellar component referred to as a thick disk by Bell (1996) (mean 
[Fe/H]$=-0.36$), Carney (2001) (mean [m/H]$\sim-0.4$ as estimated by this 
author from his Fig.5a), and Soubiran et al.\ (2003) (mean [Fe/H]$=-0.48$) 
with all these workers determining a rotational lag of 40-50 km\ s$^{-1}$. 
This component, which is expected to partially overlap the red halo both 
in [Fe/H] and age, must then be buried in the metal poor tail of the thin disk
component of Fig. 11 referred to above. Whether or not two $distinct$ stellar 
thick disk components exist cannot be resolved here although such a 
possibility was hinted at in the work of Soubiran et al. (see Norris 1999 for 
a recent report on the status of the Galactic thick disk).

We conclude that Milky Way counterparts to the components of the framework 
model do exist. It is interesting that both of the above components may have
been anticipated by the participants of the Vatican Conference on Stellar 
Populations (O'Connell, 1958). There, Oort (1958) argued for the existence of 
an `intermediate Population II' consisting of stars which formed during the 
brief contracting stage (possible identification here with our red halo). 
Likewise Baade (1958) argued for a `disk Population II' by analogy with such
a population in M31 and from the spatial distribution of what are referred to 
here as the red globular clusters which form the backbone of our thick disk.

\section{Summary and Conclusions}

Starting with a new chemical evolution model and with the goal of accounting 
for the bimodal distribution of globular clusters, 
the cosmic star formation rate density history was successfully modelled with 
a minimum number of free (i.e. non-observationally determined) parameters. The
resulting model compares favorably with the Madau plot and thus provides a 
scenario for the chemical evolution of a representative sample of the 
universe. The model was obtained by constructing a composite galaxy based on 
observational input from the Milky Way and M31. An important aspect of this 
result as presented is that it is 
independent of the effects of dust extinction. While it might be considered 
presumptuous to expect to describe the entire universe with a single galaxy, 
the summary of ancillary quantities listed in Table 3 shows that many local 
cosmological constraints are also satisfied in varying degrees.

\begin{deluxetable}{cccc}
\tablenum{3}
\tablecolumns{4}
\tablecaption{Summary of Results}
\tablehead{
\colhead{Parameter} &
\colhead{Model} & 
\colhead{Observed} &
\colhead{Reference}   
}
\startdata
 $\Omega_{\star}$ & $0.0056\pm0.0007$ & $0.0037^{+0.0019}_{-0.0013}~h^{-1}_
{0.66}$ & FHP (1998) \nl
 $\Omega_{WHIM}$ & $0.0048\pm0.0007$ & $\gtrsim0.0046~h^{-1}_{0.66}$ & Tripp 
et al.\ (2000) \nl
 $\Omega_{diffuse}$ & $0.044\pm0.002$ & $\geq0.0164~h^{-2}_{0.66}$ & 
Stocke et al.\ (2003) \nl
 $\left<[Fe/H]\right>_{WHIM}$ & $\sim-1$ & $\geq-1$ & Mulchaey (2000) \nl
 M$_{WHIM}$/M$_{\star}$ & $0.9\pm0.2$ & $>1$? & Mulchaey (2000) \nl
 M$_{\star,disk}$/M$_{\star,spheroid}$ & $0.9\pm0.2$ & $\sim1$ & Schechter \&
Dressler (1987) \nl
 $\rho(M_{t})/(\Omega_{m}\rho_{crit})$ & $0.04\pm 0.01$ & $\geq0.04$ & 
Mulchaey (2000) \nl
 $\rho_{\bullet}$(M$_{\odot}$/Mpc$^{3}$) & $2.6\pm0.5\times10^{5}$  & $2.8\pm
0.4\times10^{5}~h^{2}_{0.66}$ &
Yu \& Tremaine (2002) \nl

\enddata

\end{deluxetable}

A physical framework which incorporates the above 
results is presented. The picture involves a large scale anisotropic collapse 
of quiescently evolving, isolated protogalactic clumps. During the first 
collapse (transverse to the rotation axis), only the low angular momentum 
clumps collide resulting in the formation of the blue clusters and a reservoir
of low angular momentum gas which later is turned into bulge stars. As a 
result of the second collapse (along the rotation axis), red clusters are 
formed which become the backbone of a thick disk component. The stars already
formed in the clumps before they collide pass through and back into the halo. 
An important prediction of the model is that the Galactic halo should contain 
a relatively metal rich ([Fe/H]$_{peak}\sim-0.8$) population of 
stars with disk-like angular momentum. Galactic counterparts of this red halo 
component and model thick disk are identified and are predicted to partially 
overlap in metal abundance, age and spatial distribution. This overlap might 
explain why the properties of the `thick disk' have been so difficult to 
define up to now (e.g. Norris, 1999).

\acknowledgments 

The author wishes to thank Drs Ray Carlberg, Chris Pritchet and Sidney 
van den Bergh for commenting on earlier versions of this paper. He also wishes 
to acknowledge financial support from an NSERC (Canada) discovery grant.

\clearpage

%\begin{figure}
%\plotone{fg1.eps}
\figcaption{An illustration of the results from the new chemical evolution
model assuming a hypothetical group of globular 
clusters with Gaussian metallicity distribution parameters logZ$_{c}/
Z_{\odot}=-1.0$ and $\sigma=0.3$. Shown are the cumulative distributions of
stars, M$_{s}$ (solid line), of gas, M$_{g}$ (dashed line), of recyclable gas 
lost, M$_{ml}$ (dot-dashed line), and of mass lost due to star formation,
M$_{WHIM}$ (dotted line). Lowering logZ$_{c}$/Z$_{\odot}$ leads to a decrease 
in the ratio of stars formed to gas lost and vice versa.}
%\end{figure}

%\begin{figure}
%\plotone{fg2.eps}
\figcaption{The observed [Fe/H]-frequency histogram of M31 halo stars from
Durrell et al.\ (2001) on which is superposed the results from the chemical 
evolution model normalized to the same area. The model parameters are logZ$_
{c,red}/Z_{\odot}=-0.55$, $\sigma_{red}=0.2$, logZ$_{c,blue}/$Z$_{\odot}=-1.5$,
 $\sigma_{blue}=0.3$, M$_{t,red}/$M$_{t,blue}=2.43$, p$=0.013$, logZ$_{0}/$Z$
_{\odot}=-4.0$. Also shown are the contributions from the blue component 
(dotted line), the red component (dashed line), and the sum (solid line).}
%\end{figure}

%\begin{figure}
%\plotone{fg3.eps}
\figcaption{The age-metallicity relations for Galactic globular 
clusters. Open circles-data from VandenBerg (2000), and Salaris \& Weiss 
(2002) combined as described in the text. The solid line is the adopted 
relation for blue clusters (equations 11 \& 13), the 
dot-dashed line is the adopted relation for the red clusters (equations 11 
\& 12),
and the dashed line is the adopted relation for the bulge component (equations
11 \& 14). A representative error bar is also shown.}
%\end{figure}

%\begin{figure}
%\plotone{fg4.eps}
\figcaption{Star formation rates as a function of lookback time derived for 
each component. Shown are the blue spheroidal component (dotted line), the red 
spheroidal component (solid line), the bulge (short dashes), and the disk 
(long dashes).}
%\end{figure}
\figcaption{The Madau plot (star formation rate density versus redshift). The
solid line is the the model. The observational data (with error bars) 
are as follows: Flores et al.\ (1999) (open triangle), Treyer et al.\ (1998) 
(closed pentagon), Wilson et al.\ (2002) (closed circles), Hughes et al.\
(1998) (open circle), Giavalisco et al.\ (2003) (open squares), Steidel et al.\
(1999) (closed triangles), and Gwyn (2001) (closed diamonds). Adjustment of 
the data for cosmology and dust extinction follows from Somerville et al.\ 
(2001) for the first four data sets and from Adelberger \& Steidel (2000) 
for the last three.}

\figcaption{The age-metallicity relations transformed to the [Fe/H]-redshift 
plane on which is superposed observational data for damped Lyman $\alpha$ 
systems. Closed circles-[Zn/H] determinations discussed by Kulkarni \& Fall
(2002), closed triangles-HI column 
density weighted values of [M/H] from Prochaska et al.\ (2003). Both plotted 
quantities are [Fe/H] surrogates which account for depletion onto interstellar 
grains. The heavy lines indicate the region where the gas remaining is greater
than 1\% of its initial value. Solid line-the blue cluster relation, and 
dot-dashed line-the red cluster relation.}

\clearpage

\begin{figure}
\plotone{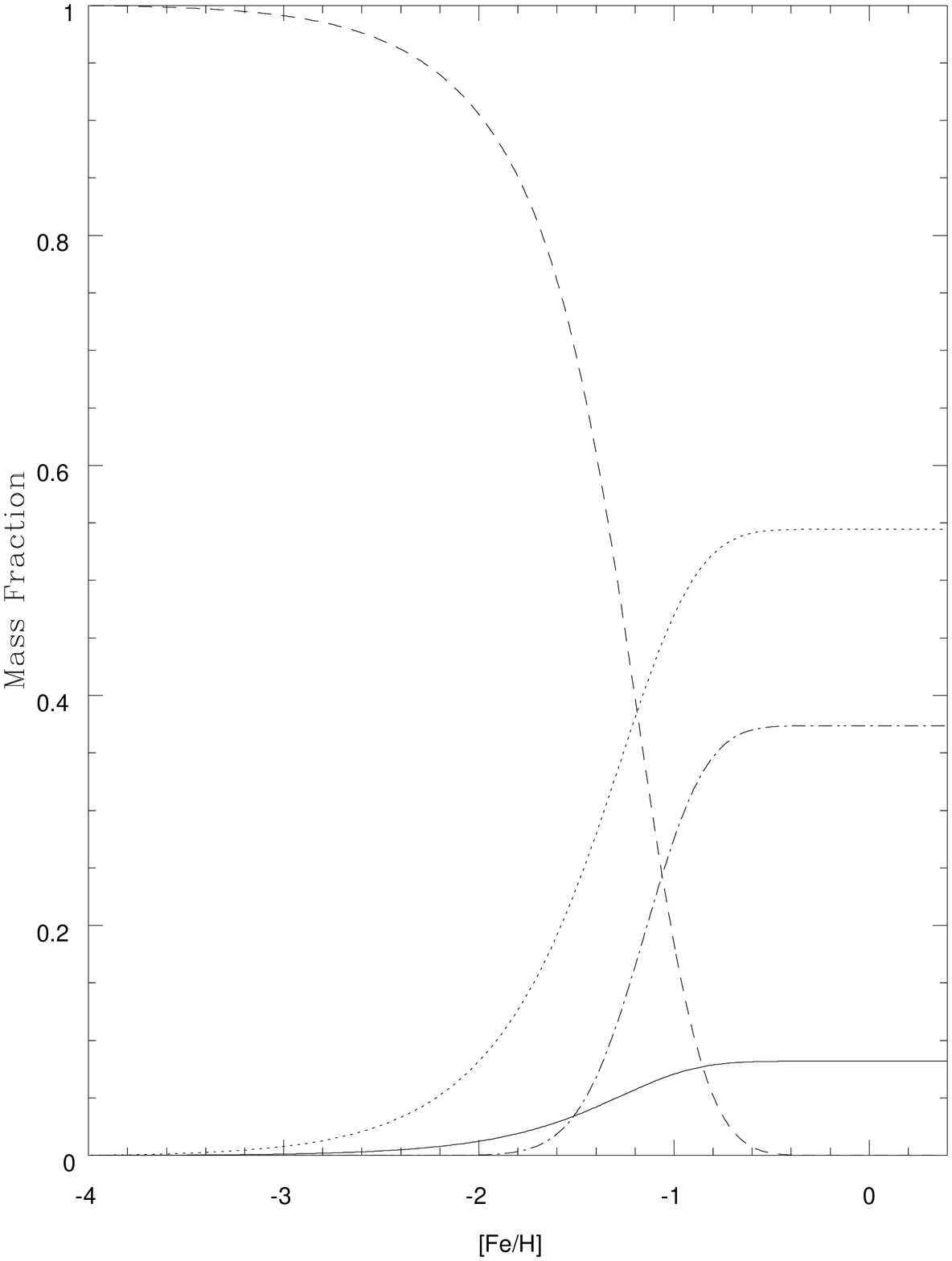}
\end{figure}

\clearpage

\begin{figure}
\plotone{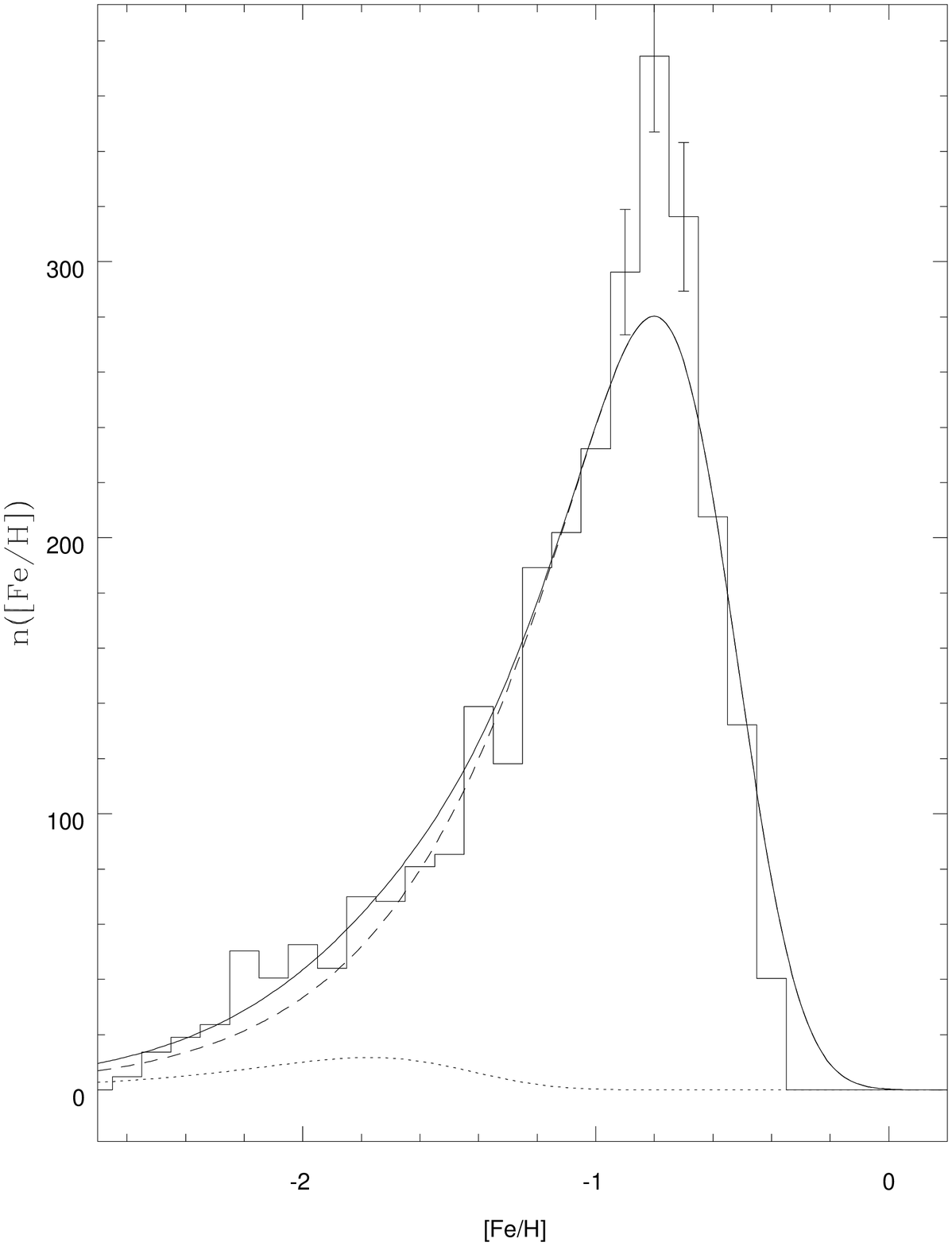}
\end{figure}

\clearpage

\begin{figure}
\plotone{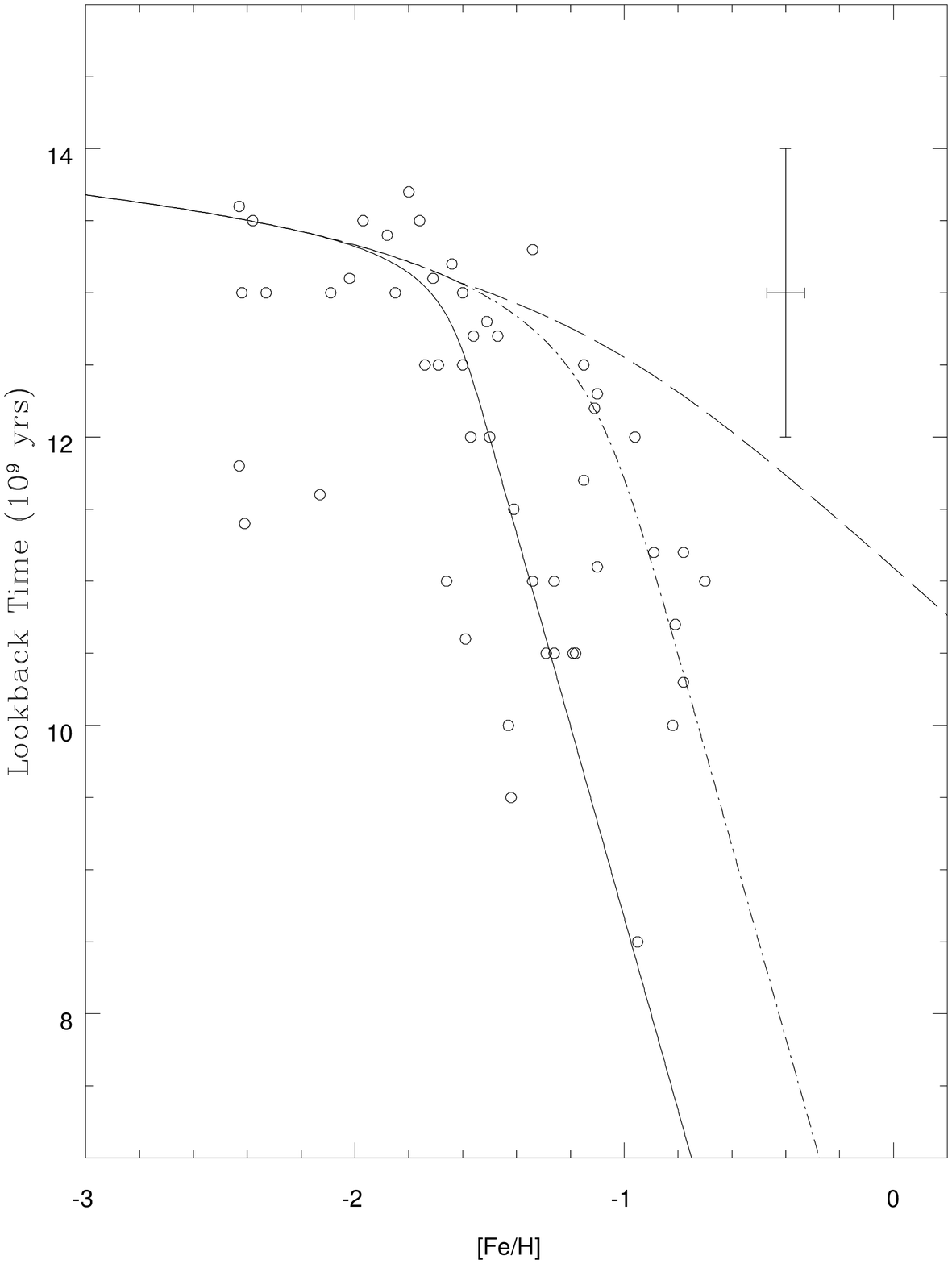}
\end{figure}

\clearpage

\begin{figure}
\plotone{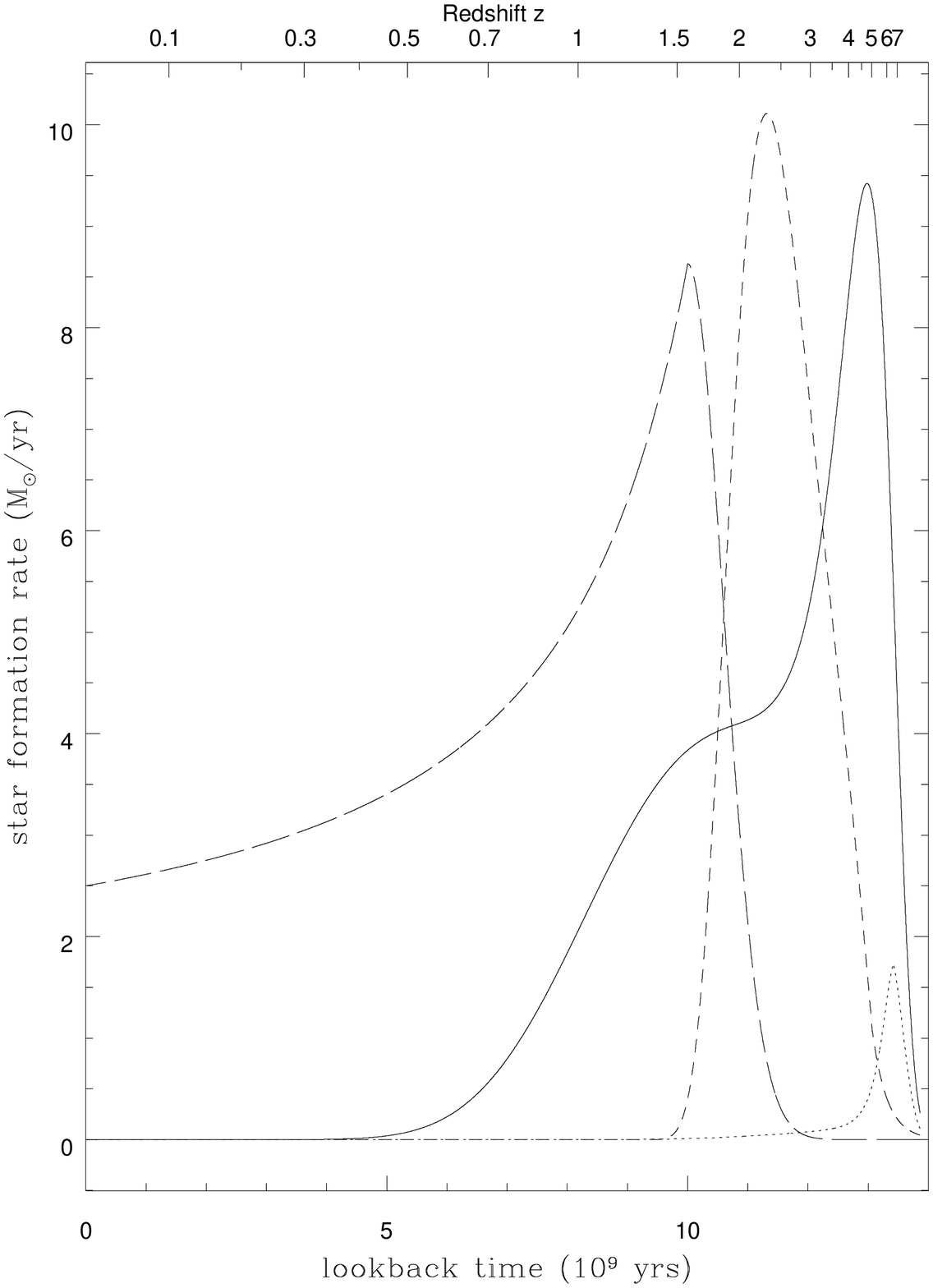}
\end{figure}

\clearpage

\begin{figure}
\plotone{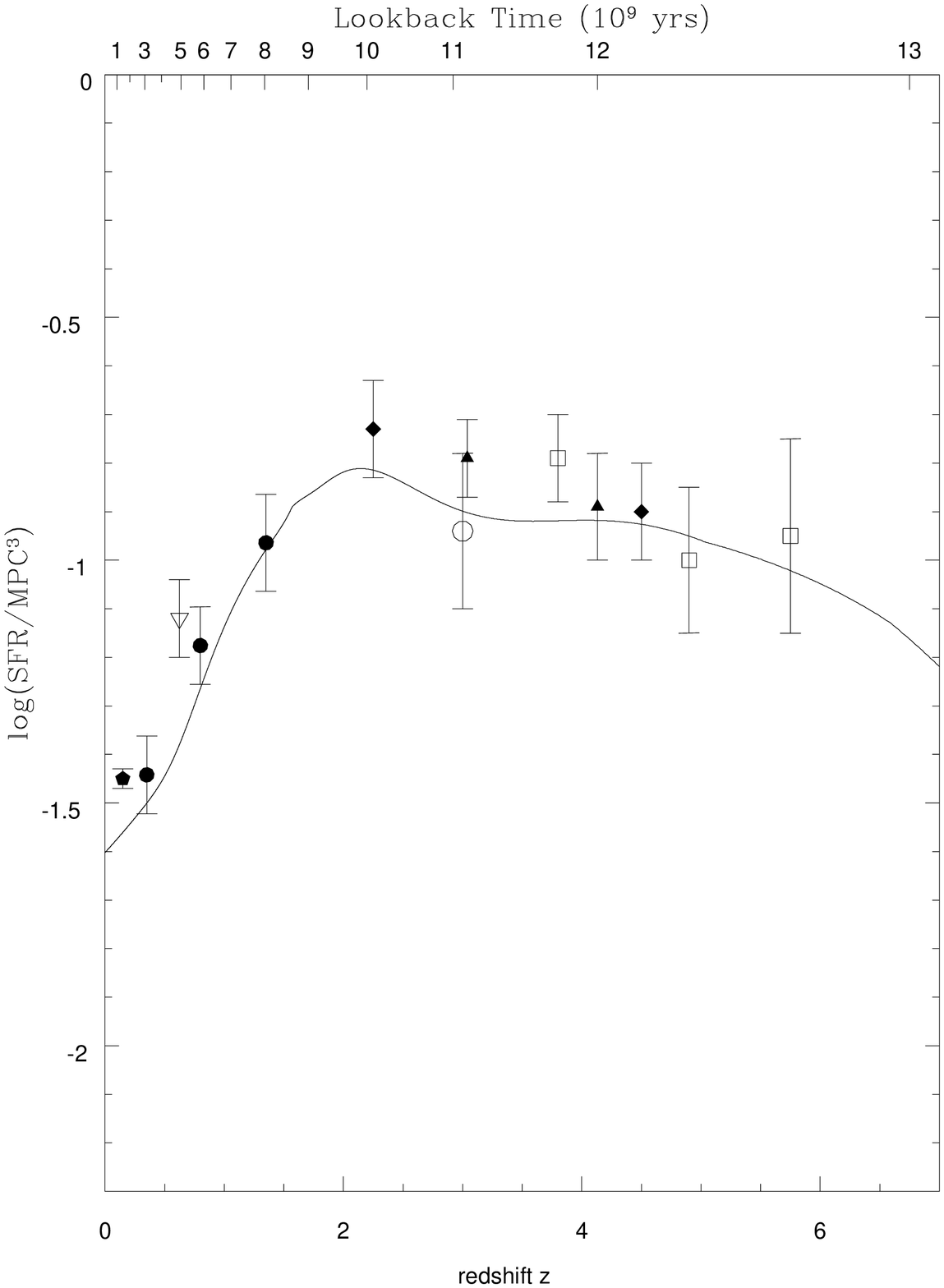}
\end{figure}

\clearpage

\begin{figure}
\plotone{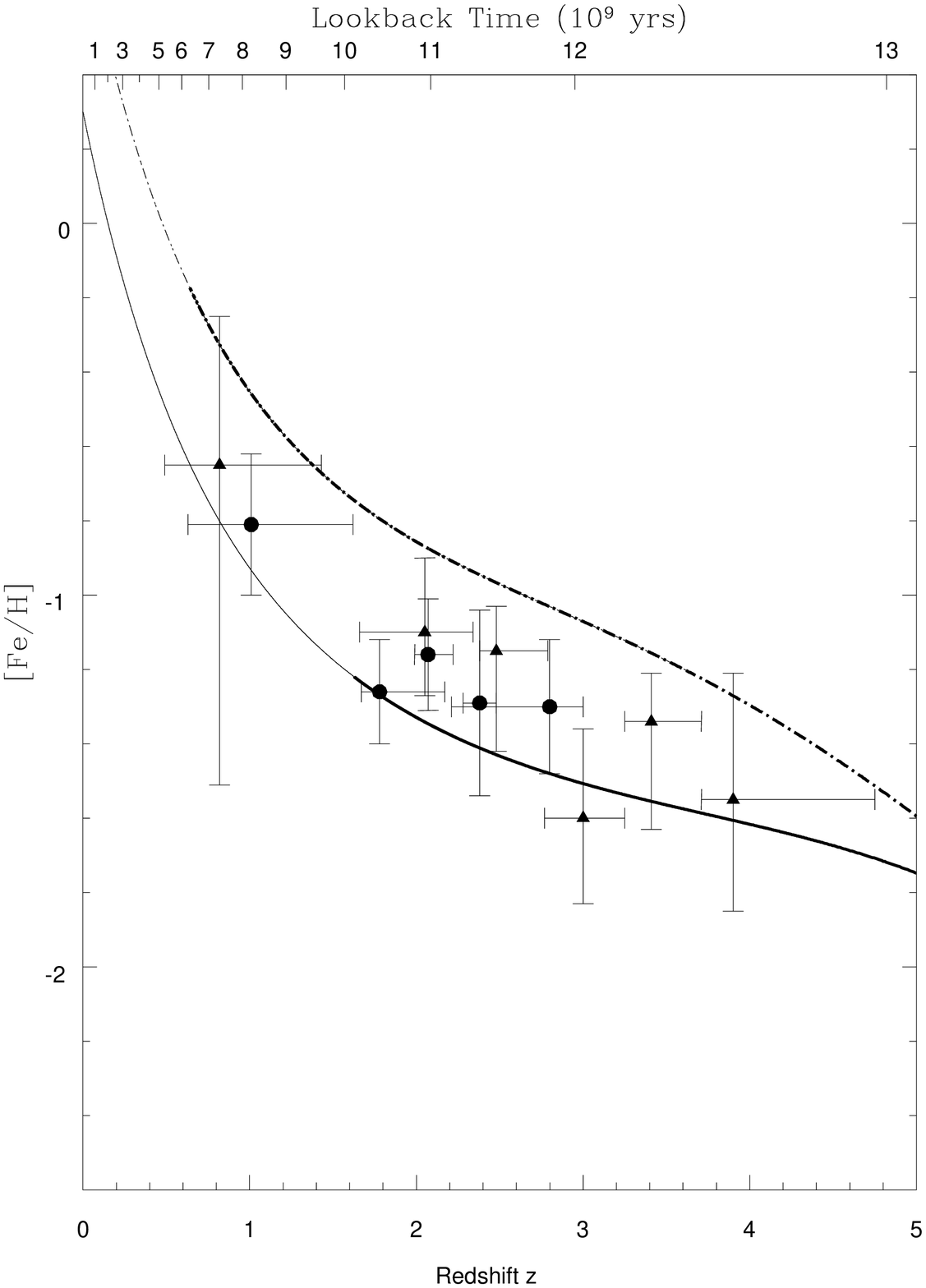}
\end{figure}

\end{document}